\documentstyle[12pt]{article}
\begin{document}
%
%
\title{
Entropy per Baryon in Strong Coupling QCD
 }
\author{
UCT-TP-223/94
\\
hep-th/9501019
\\
\vspace{0.25in}
\\
Neven Bili\'c\thanks{On leave
 of absence from Rudjer Bo\v{s}kovi\'c Institute,
 Zagreb, Croatia}\,
 and Jean Cleymans \\
Department of Physics, University of Cape Town \\ Rondebosch,
South Africa
}
\date{\today}
--
\maketitle
\begin{abstract}
The entropy per baryon is studied in the strong coupling
large dimension $d$ limit
of lattice QCD with staggered fermions. The
 partition function is calculated for
 non-zero  chemical potential and temperature using the
$1/d$ expansion.
It is found that the entropy per baryon ratio is
almost continuous across the transition from the
the quark-gluon to the hadronic phase. The relevance of these
results for heavy ion collisions is discussed.
\end{abstract}

%
%
%
%
%

Relativistic heavy ion collisions test the thermodynamic
properties of QCD.
It is well known that
lattice gauge theory is the most suitable method for
studying  these properties  theoretically,
however Monte Carlo simulations at non-zero chemical potential
$\mu$ are notoriously difficult \cite{bar86,goc88,kar89a},
 and it is necessary to investigate separately analytical methods
based on the strong-coupling,
1/$d$ expansions of QCD
at both
$T\neq 0$ and
$\mu\neq 0$.
This is the subject matter of the present investigation.
This method has been successfully applied to the  calculation of the
hadronic spectrum \cite{jol84} and
 leads to a
quantitative description of the phase diagram in the
temperature and chemical potential plane
\cite{bar86,dam86,bil88,bil92a,bil92b}.
  It yields a second order phase transition
at vanishing
$\mu$  and
predicts a first order transition for any non-zero value of $\mu$.
However, unlike the weak coupling regime, where the dimensionless
lattice parameters are related to the lattice scale via
the renormalization group equation, such a relation does not exist
in the strong coupling regime. Consequently, a physical interpretation
of the strong coupling results is not straightforward.
In this paper we propose a natural solution to this problem.
We calculate the thermodynamic properties of QCD
and contrast the results to those
obtained in the hadronic gas model.
The difference in results is very remarkable.
In the ideal gas approximation the entropy to baryon ratio (S/B) is
almost always larger in a quark-gluon plasma than in a hadronic
resonance gas, if we compare them at the same temperature and
chemical potential\cite{greiner,lee88,leonidov}. Our results provide
no support for the contention that experimental results on the charge
asymmetry ratio show evidence for quark-gluon plasma formation
\cite{lat94}.

As is usual in these considerations,
we work on an asymmetric lattice with $N_t$ points in the
time and $N_s$ in the $d$ space directions with
an anisotropy  $\gamma$.
 The finite-temperature behavior
can be studied in the infinite volume limit
($N_s \rightarrow \infty$) as a function of
$\gamma$ and $N_t$.
 In the naive
continuum limit $\gamma$ can be interpreted as a ratio
of the space and time lattice spacing, $a_s/a_t$.
As we shall shortly demonstrate, this naive relation
does not hold in the strong coupling regime. We shall
propose instead a new relation based on  general
physical considerations.

We start from the effective
partition function derived at infinite coupling
and in the large-$d$ limit \cite{fal86}
\begin{equation}\label{eq1}
Z_{\infty}=\int
[d\lambda]
\exp\{-\frac{N_c}{d}\sum_{x,y}
(\lambda_x-m) V^{-1}_{x,y} (\lambda_y-m)\}
\prod_{\vec x}\int dU_{\vec x} {\rm det} D(U_{\vec x}),
\end{equation}
with
\begin{equation}\label{eq2}
V_{x,y}=\frac{1}{2d}\sum_{k=1}^d
(\delta_{y,x+\hat{k}}
+\delta_{y,x-\hat{k}}).
\end{equation}
$D(U)$ is an $N_tN_c\times N_tN_c$ matrix
\begin{equation}\label{eq3}
D(U)=
\left(\begin{array}{ccccccc}
u_1{\bf 1}& a_1{\bf 1} &       0       &.&.&.& b_{N_t}U^{\dag}  \\
 -b_1{\bf 1}& u_2{\bf 1}& a_2{\bf 1}      & & & & 0              \\
  0      & -b_2{\bf 1}         &u_3{\bf 1}& & & & 0              \\
  .            &               &               &.& & &
  \\
  .            &               &               & &.& &
  \\
  .            &               &               & & &.&
  \\
   -a_{N_t}U   &        0      &        0     & & & &u_{N_t}{\bf 1}
\end{array} \right)
\end{equation}
with $u_i=2(\lambda_i+m)/\gamma$.
In the final expressions
the parameters $a_x$ and $b_x$ at each space-time point $x$
are set equal to
the constants
$a=e^{\mu}$ and $b=e^{-\mu}$
which are
 related to the chemical potential.
The explicit $\vec x$-dependence of the
$a_i$, $b_i$ and $\lambda_i$
in  (\ref{eq3}) has been omitted.
The integration over $dU$ can be performed by exploiting the
appropriate SU($N_c$) group integral. Details
 are given in appendices  of refs.~\cite{bil92b,fal86}.

The $1/d$ expansion begins with a saddle-point approximation of the
$\lambda$-integral, or equivalently by minimizing the
mean-field free energy
\begin{equation}\label{eq4}
F=\frac{N_tN_c}{d}\lambda^2-\ln Z_0,
\end{equation}
with $Z_0$ =
$\int dU {\rm det} D(U)$
being a ``zero-dimensional"
partition function.
We improve infinite-coupling mean-field
calculations
 by replacing the
free energy (\ref{eq4}) by
\begin{equation}\label{eq6}
F \rightarrow F -\sum_{i=1}^{5}\delta F_i,
\end{equation}
with corrections $\delta F_i$
 of order $1/g^2$ or $1/d$ \cite{bil92a,bil92b}.

Provided we know how to fix the scale $a_s$ and $a_t$
we can relate the physical
thermodynamical observables
to the corresponding lattice quantities
derived from the free energy (\ref{eq4},\ref{eq6})
\begin{eqnarray}
\label{eq7}
P=\frac{1}{a_ta_s^3} P_{latt}~~~&,&~~~
n_B=\frac{1}{a_s^3} n_{latt}\;, \nonumber \\
{\cal E}=\frac{1}{a_ta_s^3} {\cal E}_{latt}~~~&,&~~~
s=\frac{1}{a_s^3} s_{latt}\;,
\end{eqnarray}
where
\begin{eqnarray}
\label{eq8}
P_{latt} &=&-\frac{F}{N_t}-P_{vac}
\:\: ,\;\;
n_{latt}=-\frac{1}{N_t N_c}\frac{\partial F}{\partial \mu},
\nonumber \\
{\cal E}_{latt}
&=&\left.
\frac{\partial F}{\partial N_t} \right|_{\mu N_t}
-{\cal E}_{vac}\;,
 \\
s_{latt} &=& N_t (P_{latt}+{\cal E}_{latt}
-\mu N_c n_{latt}).
 \nonumber
\end{eqnarray}
 $P_{vac}$ and ${\cal E}_{vac}$ are the vacuum pressure
 and energy density calculated in the limit
 $N_t\rightarrow \infty$.
 $n_B$ and $s$ refer to the baryon density and
 the entropy density respectively. The entropy per
 baryon ratio $S/B$ is then given by $s/n_B$.

At $\mu=0$ the mean-field calculation predicts a second-order chiral
phase transition. Including $1/d$ and $1/g^2$ corrections
yields the critical anisotropy \cite{bil92b}:
\begin{equation}\label{eq9}
\gamma_c^2=
\gamma_0^2-
 \frac{\gamma_c}{g^2 N_c}(2d-\frac{N_c^2-1}{N_c}).
\end{equation}
where
the critical anisotropy at infinite coupling
takes on the value
\begin{equation}\label{eq10}
  \gamma_0^2=\frac{d(N_c+1)(N_c+2)}{6(N_c+3)}N_t .
\end{equation}

The anisotropy $\gamma$ is a measure of the temperature.
 However, unlike the weak coupling regime,
where $\gamma$ can directly be related
to the spatial and temporal lattice spacings
\cite{kar89b} allowing one
to establish the connection with
 the temperature $T=\gamma/N_t a_s$, such a direct relation
does not exist in the strong coupling limit.
In other words, the {\em physical} anisotropy of the lattice
$a_s/a_t$  is
not necessarily equal to $\gamma$.
If one
insists on using this relation in the strong coupling
regime
the critical temperature,
 owing to (\ref{eq9},\ref{eq10}),
 would strongly depend on $N_t$ which is physically unacceptable.
We must, therefore, alter the naive relation between
 $\gamma$ and the physical anisotropy (or
 temperature), in such a way as to maintain the critical
temperature independent of unphysical lattice parameters,
such as $N_t$. Quite generally we may set
\begin{equation}\label{eq11}
 a_t(\gamma,g)=\frac{a_s}{f(\gamma,g)},
\end{equation}
where the function $f$ can be determined from the
 requirements:

a) $T_c$ should not depend on $N_t$,

 b)
 at $\gamma=1$ we have
 $a_t(1,g)=a_s$.
 \\
{}From the first requirement it follows that
the $N_t$ dependence of
$f(\gamma_c,g)$ must be of the form $N_t \varphi (g)$.
Furthermore, in the limit
 $g\rightarrow \infty$ owing to
 (\ref{eq10}) we must have
 $f(\gamma_0,\infty)=\gamma_0^2$
 up to a constant which may be absorbed in
 the lattice scale $a_s$.
 Thus
 $f(\gamma_c,g)=\gamma_0^2 \varphi (g) $,
 with $\varphi$ being an arbitrary function of $g$.
 From the functional relationship between
 $\gamma_c$ and $\gamma_0$ in (\ref{eq9})
 it immediately follows
\begin{equation}\label{eq12}
f(\gamma,g)=
\gamma^2+
 \frac{\gamma}{g^2 N_c}(2d-\frac{N_c^2-1}{N_c}).
\end{equation}
up to an arbitrary multiplicative function of $g$.
The second requirement then implies
\begin{equation}\label{eq13}
a_t(\gamma,g)=a_s
 \frac{f(1,g)}{f(\gamma,g)}.
\end{equation}
%
%
%

We note that
 the correction
to the critical anisotropy
 does not depend on $N_t$
 although $\gamma_0^2$ itself is proportional to $N_t$.
The temperature can now be defined in the usual way
$T=1/N_t a_t(\gamma,g)$.
 Thus, for example, at infinite coupling  we have
 $T=\gamma^2/N_t a_s$ in
 contrast to the standard $T=\gamma/N_t a_s$.
The physical chemical potential is related to the
dimensionless lattice parameter $\mu$ as usual
$\mu_{phys}=\mu a_t(\gamma,g)$.

%
The scale $a_s$ in (\ref{eq13})
remains completely arbitrary.
 Unlike in the weak coupling regime,
where the scale can  be related to
the coupling constant via the renormalization group
equation, such a direct relation does not
exist in the strong coupling regime.
One can, nevertheless, fix the scale by
fixing the value of some physical observable, such as
$T_c$ or $\mu_c$, at a chosen value of $g$.

In Fig \ref{fig1} we show the critical temperature
 as a function of $6/g^2$ for a fixed
$\mu_{phys}$.
The reason for
 a slight $N_t$ dependence
is that the numerical
evaluation of the transition temperature
includes higher powers of $1/g^2$
which are not present in
 the analytic formula
(\ref{eq9})  used to derive
(\ref{eq12}).
The scale is fixed by requiring that
the strong coupling $T_c$ at $\mu=0$ and
$6/g^2=5$ coincides with the Monte Carlo
value $T_c=130$ MeV \cite{alt93}.

The critical entropy per baryon
in the hadronic phase
increases rapidly with decreasing $g$
(Fig \ref{fig2}).
If we fix the coupling at the point where
the strong-coupling curve crosses the
experimental region (hatched area)
\cite{let93} we can plot the
entropy per baryon as a function of temperature
on both sides of the transition point and compare
this with the quark-gluon plasma and hadronic gas
 model (Fig \ref{fig3}).
 In contrast to this simple
phenomenological model, strong coupling lattice
thermodynamics
predicts a continuous $s/n_B$ across the phase transition.
 In the infinite coupling
  limit ($6/g^2=0$) the entropy per baryon in the
 chiral-broken (hadronic) phase is greater than
 in the symmetric (quark-gluon) phase.
%
%
%

Preliminary data from the
emulsion experiment EMU05
cited in \cite{let93}
show that in S-Pb collisions at 200 GeV the charge asymmetry ratio
\begin{equation}\label{eq14}
\frac{N^+-N^-}{N^++N^-}
\end{equation}
is approximately 0.085$\pm$0.01 in the central rapidity
region.
To relate this to the entropy per baryon ratio one can proceed as
follows.
Assuming that this region is symmetric in isospin (for S-S collisions
this would be exact),
the numerator in (\ref{eq14}) equals the baryon number divided by 2,
assuming furthermore
that the denominator is approximately 2/3 of the total
particle number (neutral particles are not included),
the ratio becomes :
\begin{equation}\label{eq16}
\frac{N^+-N^-}{N^++N^-}\approx \frac{3B}{4N}
\end{equation}
To relate this to the entropy we use as a guess
that the entropy per particle
is 4 (as is the case of a massless Boltzmann gas).
We thus conclude that
\begin{equation}
S/B\approx 36\pm 5
\end{equation}
This estimate is compatible with the more detailed
analysis presented in\cite{let93,heinz}. This result was used above
in Fig. 2.

Contrary to the predictions of
quark-gluon plasma - hadronic gas models
 we find in the strong coupling regime
  that the entropy per baryon is almost
 continuous across the phase transition.
 Moreover, including the strong coupling corrections increases the
 entropy per baryon content in the hadronic phase
 near the phase transition to
 the order close to the estimates obtained from the
 measured charge asymmetry ratio.

In conclusion, we have calculated the thermodynamic properties of QCD
in the strong coupling large $d$ limit at finite temperature and
chemical potential.
We have introduced a new relation between the lattice anisotropy
parameter $\gamma$ and the physical anisotropy
$a_s/a_t$ allowing us to establish a physical interpretation
of the dimensionless parameters
$\gamma$ and $\mu$ in terms of
 temperature and the chemical potential.
 Our results show that the entropy per baryon
ratio $S/B$ is almost continuous across the phase transition point.
\vspace{.3in}

 {\large\bf  Acknowledgments}

\vspace{.2in}
We acknowledge useful discussions with Krzysztof Redlich.
We furthermore acknowledge financial support from the Foundation for
Research and development (FRD), Pretoria.

\begin{figure}[p]
\caption{Critical temperature at fixed $\mu=0$
and $\mu=250$ MeV for
$N_t=4$ and 6.}
\label{fig1}
\end{figure}
\begin{figure}[p]
\caption{Critical entropy per baryon in hadronic phase
at fixed $\mu=250$ MeV for
$N_t=4$ and 6.}
\label{fig2}
\end{figure}
\begin{figure}[p]
\caption{Entropy per baryon calculated
in strong coupling QCD compared to
quark-gluon plasma (QGP) and hadronic gas
model at fixed $\mu$=250 MeV.}
\label{fig3}
\end{figure}
\end{document}